# Mechanism for long-acting chimeras based on fusion with the carboxyl-terminal peptide (CTP) of human chorionic gonadotropin beta-subunit 3

J. C. Phillips

Dept. of Physics and Astronomy, Rutgers University, Piscataway, N. J., 08854

Abstract

Thermodynamic scaling explains the dramatic successes of CTP fused human growth proteins as regards lifetime in vivo and enhanced functionality compared to their wild-type analogues, like Biogen. The theory is semi-quantitative and contains no adjustable parameters. It shows how hydrophilic terminal spheres orient fused proteins in the neighborhood of a membrane surface, extending lifetimes and improving functionality.

There is growing interest in the cloning of chimeras based on fusion of a wild-type protein with the 28 amino acid (aa) carboxyl-terminal peptide CTP = 138SSSS…ILPQ165 (Uniprot P0DN86 numbering) of human chorionic gonadotropin (hCG) beta-subunit 3 (217 aa). The advantageous in vivo survival properties of wild type proteins fused to the beta fragment were discovered some time ago in connection with follicle stimulating hormones [1]. More recently biopharmaceutical attention has shifted to fused human growth protein (fHG) [2,3], where clinical tests have yielded promising results for enhanced protein survival [2,3]. Similar successes have appeared for endostatin (2fEND) by fusing double CTP [4].

Here we are concerned with the mechanism involved. Fusion is promoted by the SSSS lead, but why? Similarly, CTP overall resembles a large hydrophilic glycoprotein, which generally increases hydrophilic shielding and inhibits metabolic clearance [2]. Is CTP merely an expanded polymer (like polyethylene glycol, PEG [5], but larger), or is its action more specific?

Questions of this kind can be discussed using thermodynamic scaling, which yields epitopic profiles of protein segments of length N aa or longer (N ≥ 9) especially with regard to lengths N

~ 20 aa in the membrane thickness range.  CTP is just such a segment, and the method explains many features of allosteric interactions [6,7].  The scaling method quantifies general features of protein folding, especially those associated with hydropathic globular shaping and protein functionality.  It is especially useful in characterizing hydropathically shaped global motions that are discussed (for example) in elastic network models [8].  Its accuracy is possible because it is based on the globular hydropathic fractals discovered bioinformatically [9].

In order to reflect allometric interactions, one must average the individual aa hydropathicities over a sliding window of width W.  Nine of the first ten aa of CTP {SSSSKAPPPS} are hydrophilic, which is unusual {average $\Psi$ = 109, compared to hydroneutral, 155}.  Near their centers mucin repeats [10] are also strongly hydrophilic, and there, ~ 10% of the amino acids are again Ala or Gly (hydroneutral).   Thus we have chosen to show profiles with W = 11.

The pharmaceutical benefits of CTP fusion to GH (wild type) were studied in [2] in terms of rat weight gain (their Fig. 3).  Here the benchmark is Biogen administered daily, while GH and its CTP-fused variations were administered weekly. Biogen daily is beneficial at 17, compared to weekly GH (wt) 4, GH-CTP (6) and CTP-GH (8) improved GH, with CTP-GH slightly better than GH-CTP.  The results improved dramatically for CTP-GH-CTP (19, better weekly than Biogen daily). The dramatic synergistic improvement of CTP-GH-CTP compared to CTP-GH and GH-CTP reflects allosteric interactions.

Thermodynamic scaling explains the dramatic success of the fused CTP, GH variants (Fig.2). Broadly speaking, with regard to dynamic hydropathic distortions, GH is divided into three parts: a central part dominated by a deep hydrophilic hinge, and N- and C terminals dominated by hydrophobic peaks. As the figure caption explains, the hydrophilic CTP additions have a larger effect near the N terminal because the hydrophobic peak is larger there.

A deeper discussion explains the allosteric interactions.  Transmembrane protein segments are generally hydrophobic, and this hydrophobic interaction should continue weakly into the "frontier" layer adjacent to the membrane.  (This idea was proposed previously [7,11] to explain scaling results for other proteins.  This picture is analogous to heterogeneous catalysis, with the membrane playing the part of catalytic substrate.)  The greater weight gain and the dramatically longer lifetime of CTP-GH-CTP could result from more complete and stable  shielding its very soft central hinge ($\Psi$(aa,W = 11) touches 100 in Fig. 2) from proteolysis, itself catalyzed by proteases (Wiki). The differences between the Myoglobin hydroprofiles of freshwater and ocean fish is a simple example of the stabilizing effects of terminal hydrophobic peaks [12].

Endostatin is the C-terminal fragment 1572-1754 of Collagen alpha-1(XVIII) chain (Uniprot P39060) 1572SHRD…TASK1754. As expected for collagen, it is overall hydrophobic, especially near the C terminal (Fig. 3). This explains why [4] fused CTP twice to form a chimera with a doubly hydrophilically shielded endostatin C terminal. Unlike HG, the rest of the hydroprofile (Fig. 4) shows little structure. There is no deep central hydrophic hinge, and a weak hydrophobic N-terminal peak ($\Psi$(aa,W = 11) reaches only 176, compared to 200 for the C-terminal peak. Thus the best fused CTP structure for endostatin END-CTP-CTP is also explained immediately by the strengths of its terminal hydrophobic peaks.

Human growth hormone is often associated with prolactin (PRL, full profile, Fig. 4), and the two proteins may have a common ancestor [13]. Their overall aa similarity, as measured by BLAST, is marginal, so the question of whether a fused PRL enhancer should be developed is interesting, either as an HG adjuvant, or because low levels of prolactin could be medically important [14]. There are large mechanistic differences between the C-terminal HG and PRL [15]. Their hydropathic profiles are compared in Fig. 5; near the C terminal they are quite similar. As for the PRL N-terminal, its hydrophobic peak of ($\Psi$(aa,W = 11) reaches 189, compared to only 170 for the N-terminal peak (Fig. 4). Thermodynamic scaling predicts that an effective fPRL chimera involving CTP would be optimal the CTP-CTP-PRL configuration.

In conclusion, our key point is that hydrophilic terminal spheres (both PEG [16] and CTP) orient the fused protein in the neighborhood of a membrane surface, shielding the fused protein proteolytic regions, which could contribute to its survival and even enhance its activity with other nearby proteins (such as receptors) similarly oriented. We tested this conclusion against CTP data on fHG and fEND. If we divide each protein into three parts (central and N- and C-terminal regions), then the results agree with all available data using only the natural CTP choice W = 11 for the sliding window length; otherwise there are no adjustable parameters. Should PRL also prove to be of interest, we have made predictions there, which suggest results different from those of its companion HR protein.

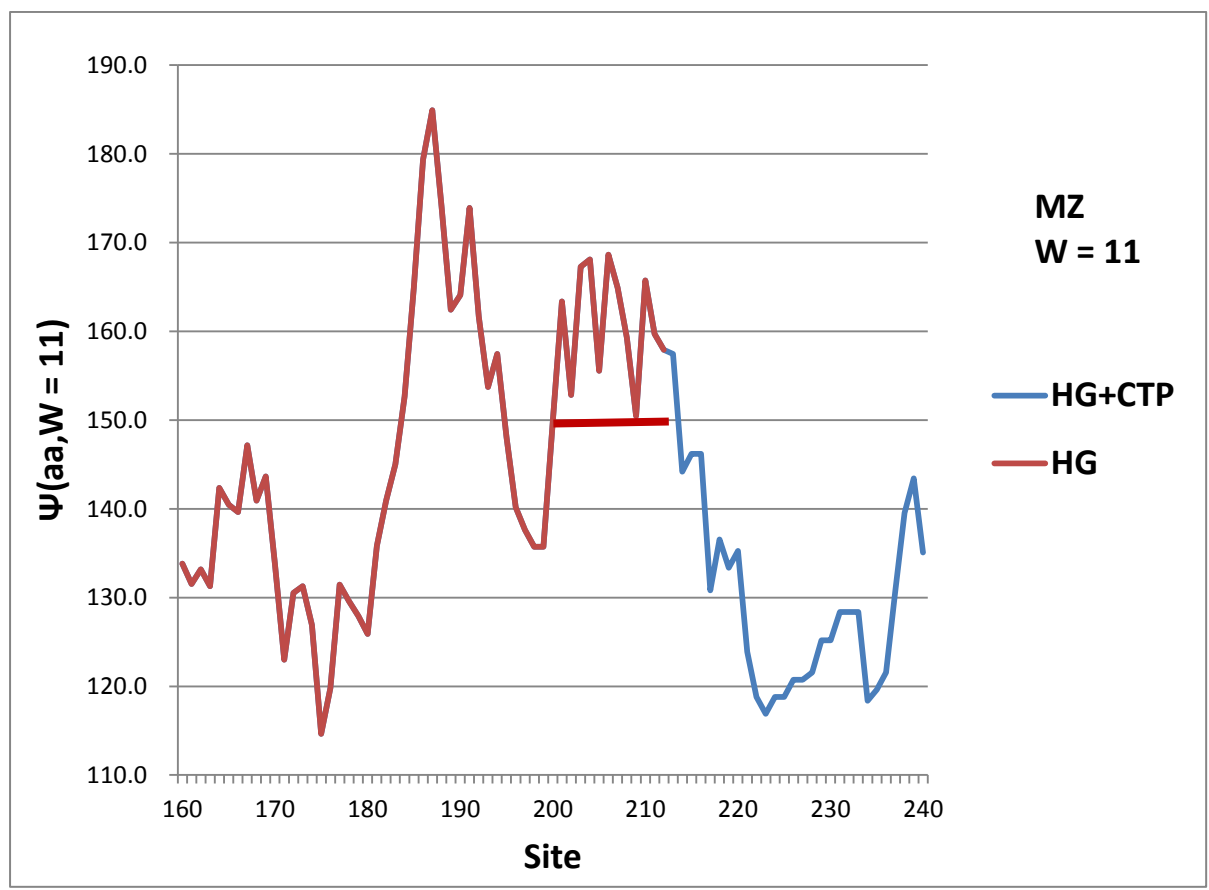

Fig. 1.  Hydropathic profiles using the MZ scale near the C terminal of HG.  It has been scaled so that large values are hydrophobic, small values hydrophilic, and hydroneutral (Ala, Gly and His) is 155.  Fusion with CTP adds a hydrophilic sphere to the hydrophobic C terminal of human growth protein (HG). The largest value of $\Psi(aa, W = 11)$ over the aa in the last HG hydrophobic peak (205-217, marked by line in figure) is 169 (well above nearly hydroneutral = 155).

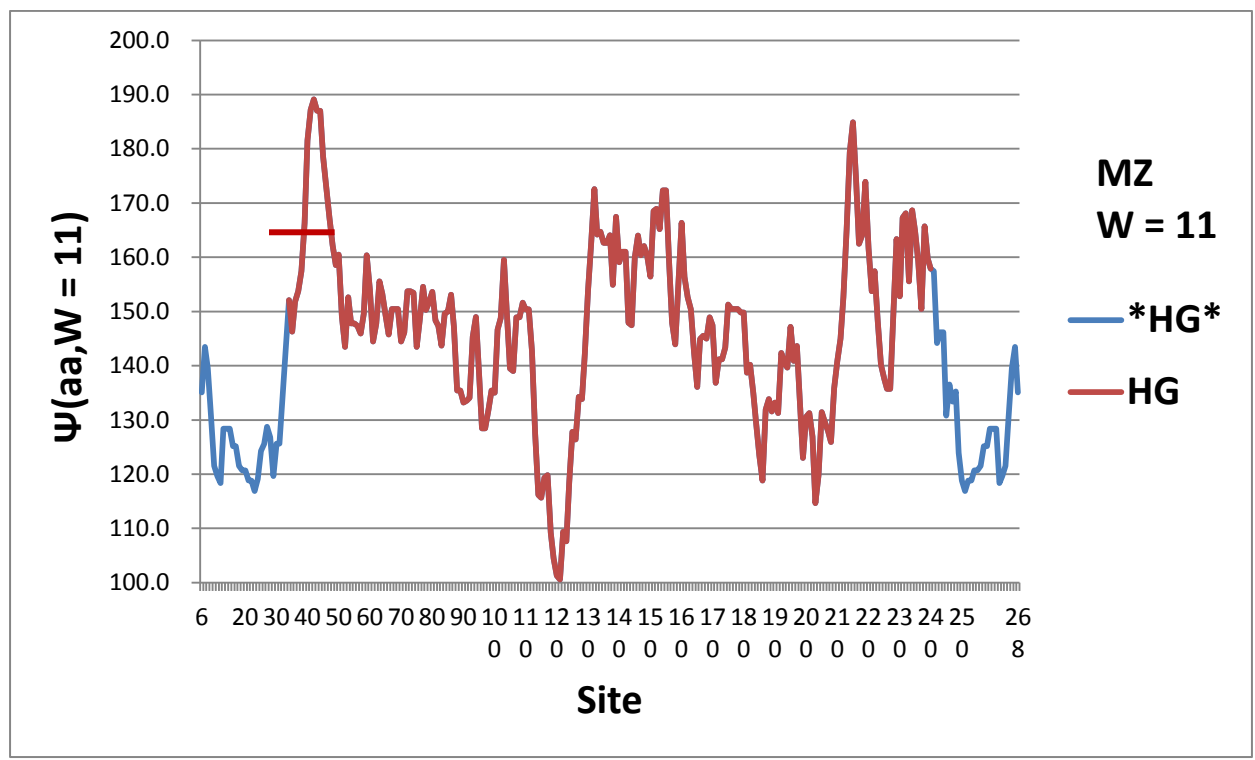

Fig. 2. The complete CTP + HG + CTP = *HG* protein, with hydrophilic CTP fused to both N and C terminals. The largest value of $\Psi(aa, W = 11)$ over the aa in the first HG N terminal hydrophobic peak (marked by line in figure) is 189 (2.5 times higher above hydroneutral = 155) than the largest value of the C-terminal peak). The site numbering includes a 28 aa offset due to the N-terminal CTP.

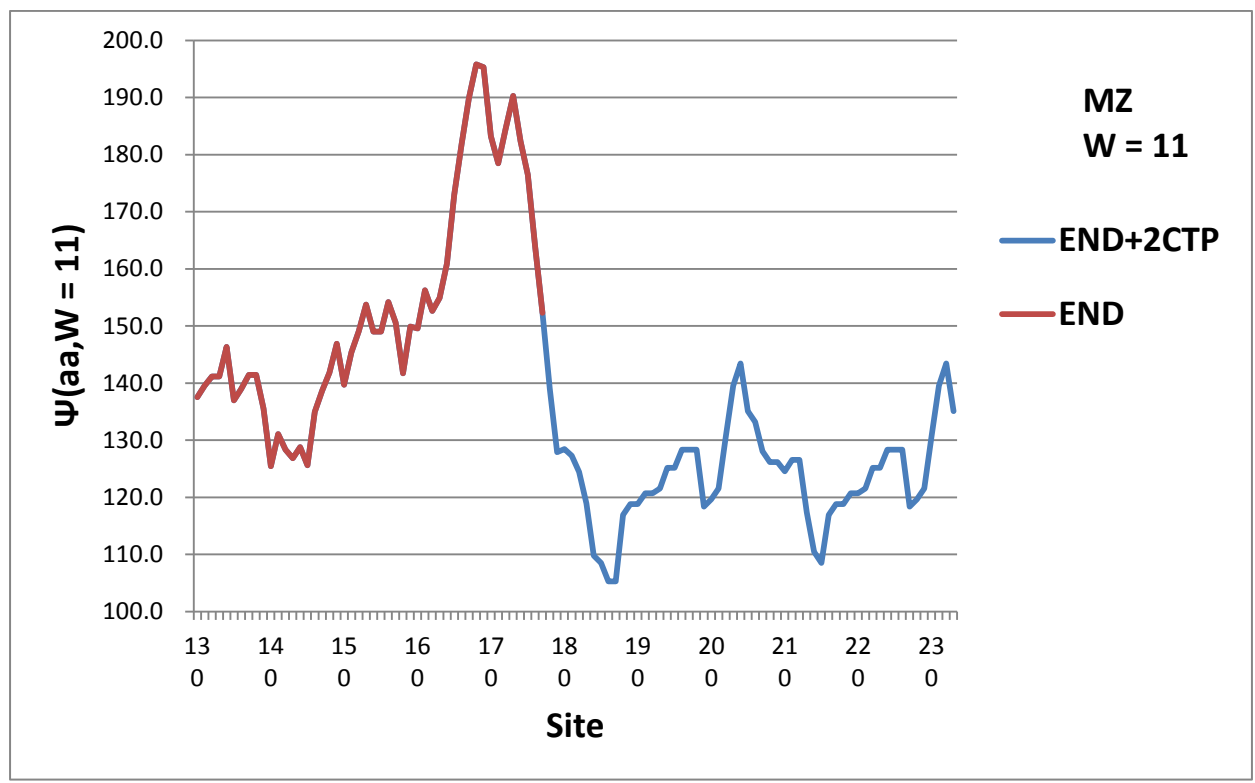

Fig. 3. Hydropathic profiles of of C terminal of endostatin using the MZ scale. Compared to HG, where the last HG hydrophobic peak averaged 158.5 (hydroneutral), here the hydropathic peak (163-178) averages 176.0 – strongly hydrophobic. Its peak value is close to 200, larger than that of the HG C terminal peak.

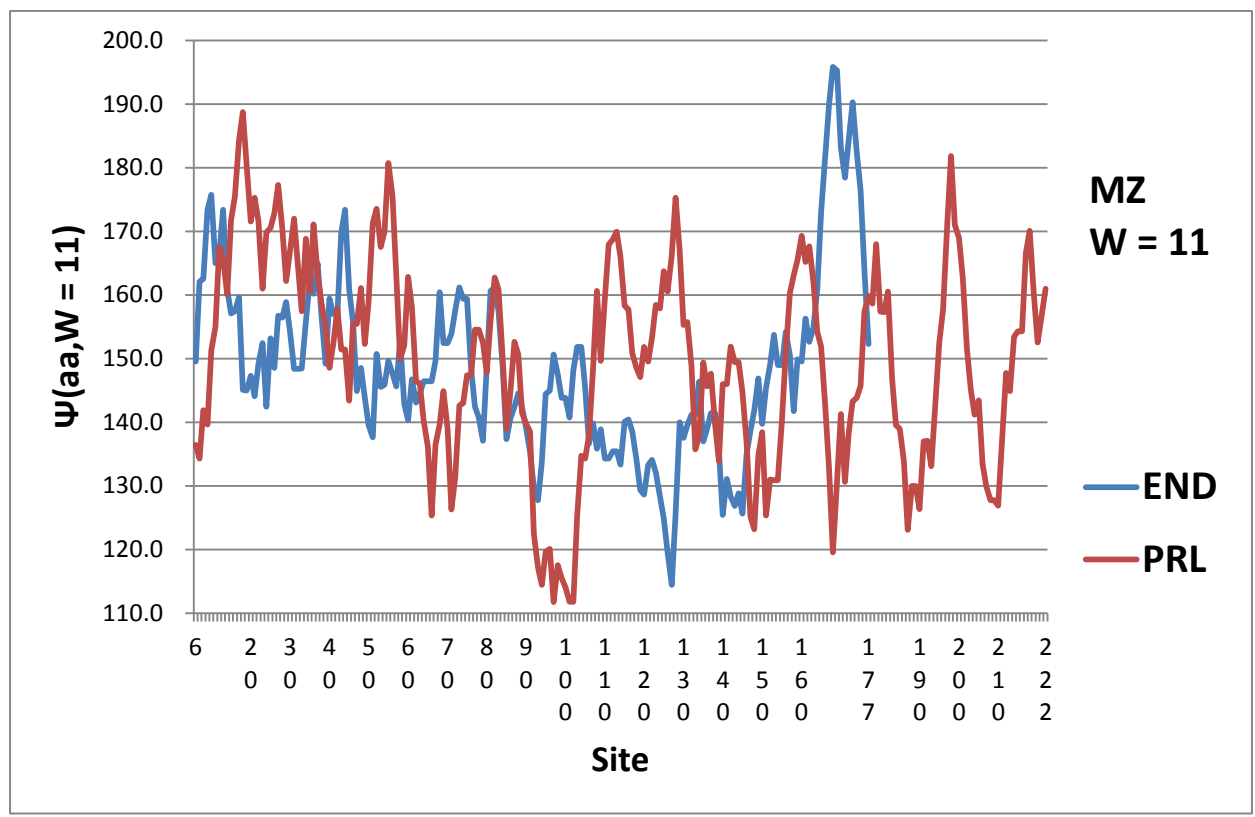

Fig. 4. The hydroprofiles of mendostatin and prolactin exhibit similar central hydrophilic hinges, but their hydrophobic peaks near their C- and N- terminals are qualitatively different.

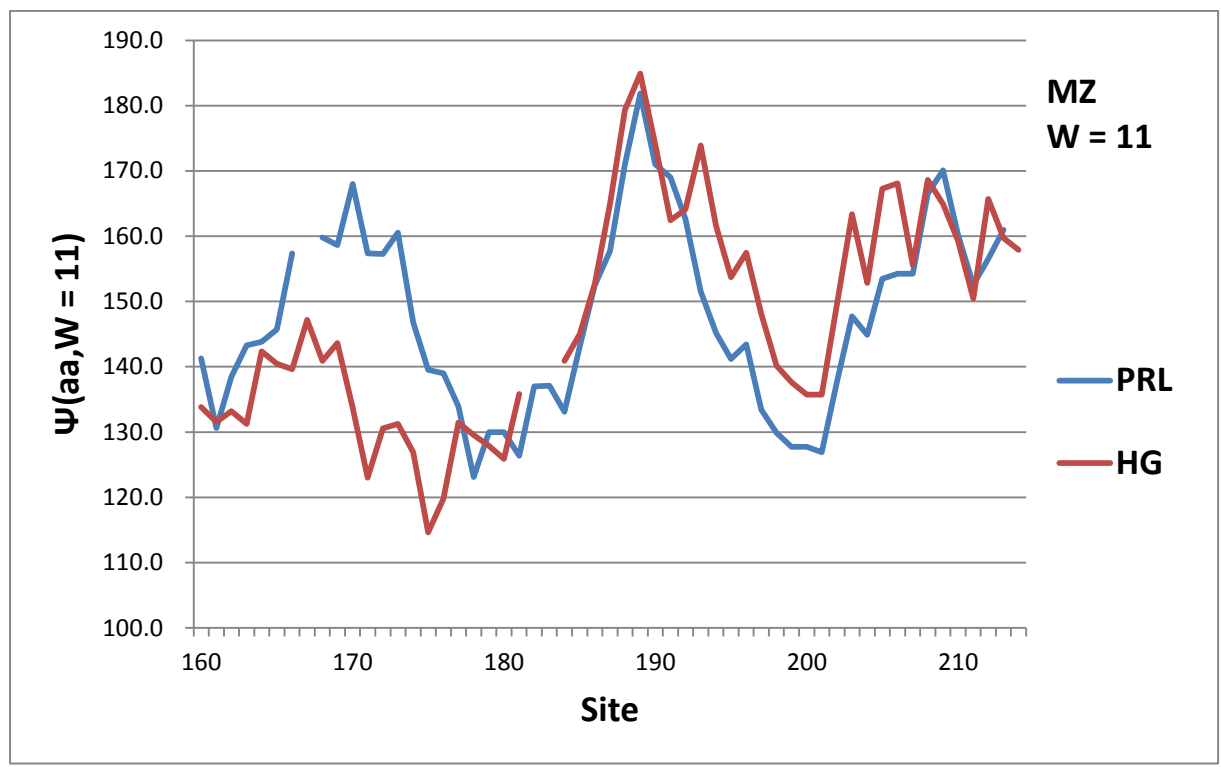

Fig. 5. Comparison of BLAST-aligned PRL and HG (Uniprot HG numbering). Above site 200 the two profiles are quite similar, suggesting the possibility of a PRL chimera similar to the HG chimera [2,3].